\documentclass{elsart}

\usepackage{amssymb}
\usepackage{graphicx}

\begin{document}

\begin{frontmatter}



\title{Structure of isomeric states in $^{66}$As and $^{67}$As}


\author[a1]{M. Hasegawa},
\author[a2,a3]{Y. Sun},
\author[a4]{K. Kaneko},
\author[a5]{T. Mizusaki}

\address[a1]{Laboratory of Physics, Fukuoka Dental College, Fukuoka 814-0193, Japan}
\address[a2]{Department of Physics and Joint Institute for Nuclear Astrophysics,
University of Notre Dame, Notre Dame, IN 46556, USA}
\address[a3]{Department of Physics, Xuzhou Normal University,
Xuzhou, Jiangsu 221009, P.R. China}
\address[a4]{Department of Physics, Kyushu Sangyo University,
Fukuoka 813-8503, Japan}
\address[a5]{Institute of Natural Sciences,
Senshu University, Tokyo 101-8425, Japan }

\begin{abstract}
Strong residual correlations between neutrons and protons in
$N\approx Z$ systems can lead to unusual structure. Using the
spherical shell model, we show that a low-excitation shape isomer
can occur in the odd-odd $N=Z$ nucleus $^{66}$As. This extends the
picture of shape coexistence beyond even-even nuclei. Furthermore,
it is demonstrated that in $^{66}$As and in the $N=Z+1$ nucleus
$^{67}$As, a new type of isomer, which we term $j$-isomer, can be
formed. The underlying mechanism for the isomerism formation is
structure change in the isomeric states, which involves either an
alignment of a neutron-proton pair from the high-$j$ intruder
orbitals, or a simultaneous occupation of these neutron and proton
high-$j$ orbitals.

\end{abstract}

\begin{keyword}
isomeric states \sep $N\approx Z$ nuclei \sep shell model

\PACS 21.60.Cs \sep 21.10.-k \sep 27.50.+e
\end{keyword}
\end{frontmatter}


Nuclear structure study near the $N=Z$ line from nickel to $A \sim
100$ nuclei is one of the current research topics in nuclear
physics. There are at least two reasons why one is interested in
these nuclei. First, this region is characterized by rapid shape
change \cite{Lister} from nucleus to nucleus, exhibiting abundance
of interesting structure. The occurrence of the rich phenomena,
with some of their origins still unknown, is unique to this mass
region. It is likely due to the occupancy of the same orbitals by
neutrons and protons, which leads to strong residual
neutron-proton correlations. Second, it is anticipated \cite{Sun}
that the unique structure may have implications in the
understanding of nucleosynthesis. It has been suggested that the
proton-rich nuclei near the $N = Z$ line are synthesized in the
rapid-proton capture process (the rp-process) \cite{rpReport}
under appropriate astrophysical conditions, such as in the X-ray
burst environment. The rp-process creates nuclei far beyond
$^{56}$Ni all the way to the heavy proton-rich regions of the
chart of nuclides. However, current network simulations for the
rp-process may contain large uncertainties because of the lack of
information on the structure near the $N=Z$ line.

Among the variety of structure discussed in this mass region,
nuclear isomer is probably the least understood one. Isomeric
states are excited metastable states, which are formed only in a
limited number of nuclei depending on the detailed shell structure
of the neutron and proton orbitals. Often discussed in the
literature are three mechanisms \cite{Nature} leading to nuclear
isomerism. It is difficult for an isomeric state to change its
shape to match the states to which it is decaying, or to change
its spin, or to change its spin orientation relative to an axis of
symmetry. These correspond to shape isomers, spin traps, and
$K$-isomers, respectively. In any of these cases, $\gamma$-decay
to the ground state is strongly hindered, either by an energy
barrier or by the selection rules of electromagnetic transition.
Therefore, isomer lifetimes can be remarkably long. As recently
studied examples in the $A=60-100$ nuclei, a $J^\pi = 0^+$ excited
state in $^{72}$Kr has been found as a shape isomer
\cite{Bouchez}, a $12^+$ state in $^{98}$Cd as a spin trap
\cite{Blazhev}, and a $9^+$ excited bandhead in $^{70}$Br,
possibly as a $K$-isomer \cite{Jenkins0}.

Experimental study of excited states in $N \approx Z$ nuclei is a
rather challenging problem. Despite the difficulties, information
about isomeric states has been gathered in recent years. The
nucleus $^{66}$As has aroused a special interest in the study of
odd-odd $N=Z$ nuclei since the discovery of two isomeric states,
the $J^\pi = 5^+$ one at 1.357 MeV and the $9^+$ one at 3.024 MeV
\cite{Grzywacz1,Grzywacz2}. In the lighter odd-odd $N=Z$ nucleus
$^{62}$Ga, an isomeric state with $3^+$ at a much lower excitation
0.817 MeV has been known \cite{Vincent,Rudolph}. It has also been
reported that the odd-mass nucleus $^{67}$As has an isomeric state
with ${9/2}^+$ \cite{Jenkins}. On the theoretical side, some of
these nuclei have been investigated by different theoretical
models, such as the shell model \cite{Vincent,Rudolph},
interacting boson model \cite{Juillet}, and a deformed shell model
\cite{Sahu}. However, the question of why the above mentioned
states become isomeric and what the nature of the isomerism is has
not been thoroughly addressed.

In this Letter, we investigate the structure of isomeric states in
odd-odd $N=Z$ nucleus $^{66}$As and in adjacent odd-mass nucleus
$^{67}$As. The calculation is performed by using the spherical
shell model. Our discussion on the results will focus on those
states that have notably small electromagnetic transition
probabilities to the low-lying states. Analysis shows that there
are essentially two different types of isomers entering into
discussion. Due to the fact that prolate and oblate shapes can
coexist at the low excitation region, a shape isomer is predicted
in $^{66}$As. The other type of isomer is related to a suppressed
decay between structures based on the high-$j$ $g_{9/2}$ intruder
orbital and the $pf$-shell configurations.

We start with a general form of the extended $P+QQ$ Hamiltonian
which is composed of the single-particle energies, $T=0$ monopole
field, monopole corrections, pairing forces with $J=0$ and $J=2$,
quadrupole-quadrupole ($QQ$) force and octupole-octupole ($OO$)
force \cite{Hase0,Kaneko0}:
\begin{eqnarray}
 H & = & H_{\rm sp} + H^{T=0}_{\pi \nu} + H_{\rm mc}
        + H_{P_0} + H_{P_2} + H_{QQ} + H_{OO}  \nonumber \\
   & = & \sum_{\alpha} \varepsilon_a c_\alpha^\dag c_\alpha
       - k^0 \sum_{a \leq b} \sum_{JM} A^\dagger_{JM00}(ab) A_{JM00}(ab)
            \nonumber \\
   & + & \sum_{a \leq b} \sum_{T} \Delta k_{\rm mc}^T(ab) \sum_{JMK}
               A^\dagger_{JMTK}(ab) A_{JMTK}(ab)         \nonumber \\
   & - & \sum_{J=0,2} \frac{1}{2} g_J \sum_{M\kappa} P^\dag_{JM1\kappa} P_{JM1\kappa}           \nonumber \\
   & - & \frac{1}{2} \chi_2 \sum_M :Q^\dag_{2M} Q_{2M}:
         - \frac{1}{2} \chi_3 \sum_M :O^\dag_{3M} O_{3M}:.
          \label{eq:0}
\end{eqnarray}
This isospin invariant Hamiltonian is diagonalized in the chosen
model space based on a spherical basis \cite{Mizusaki}. In the
present work, we employ the $pf_{5/2}g_{9/2}$ model space. This
shell model has recently proven to be rather successful in
describing nuclear shapes, energy levels, and the band-crossing
phenomenon in Ge and Zn isotopes, as well as in $^{68}$Se
\cite{Hase1,Kaneko,Hase2}. For the interaction strengths, we adopt
the same ($A$-dependent) parameters as those employed in Ref.
\cite{Hase2}, except that we add the following monopole terms:
\begin{eqnarray}
 \Delta k_{\rm mc}^{T=0}(a,g_{9/2}) = -0.18 \mbox{ MeV},
 \quad a=p_{3/2}, f_{5/2}, p_{1/2}.   \label{eq:1}
\end{eqnarray}
These additional monopole terms have an effect of lowering the
$g_{9/2}$ orbital, which is needed for obtaining correct positions
of the $9_1^+$ state and the higher spin states relative to the
$7_1^+$ state in $^{66}$As. Such additional monopole terms are
found necessary also for other odd-odd $N=Z$ nuclei in our
calculations, though there remains a question why odd-odd $N=Z$
nuclei require these terms. The $T=0$ monopole field $H^{T=0}_{\pi
\nu}$ affects significantly the relative energy between the $T=1$
and $T=0$ states in odd-odd $N=Z$ nuclei \cite{Hase0}. We have
determined the strength $k^0$ to be $1.2 (64/A)$, which roughly
reproduces the excitation energies of the $7_1^+$ and $9_1^+$
states in $^{66}$As. The effective charges used in the calculation
are $e_{\rm eff}^\pi=1.5e$ and $e_{\rm eff}^\nu=0.5e$.

\begin{figure}[t]
\includegraphics[width=8cm,height=7.5cm]{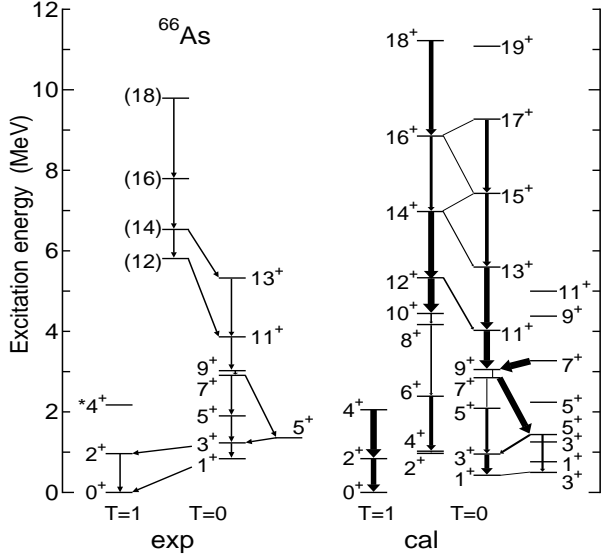}
  \caption{Experimental and calculated energy levels in $^{66}$As.
           The arrows between the experimental states (exp) indicate the
           observed electromagnetic transitions in Ref. \cite{Grzywacz2},
           and widths of the arrows between the calculated states (cal)
           denote relative values of the calculated $B(E2)$.
           The experimental energy of the $T=1$, $4_1^+$ state
           is taken from $^{66}$Ge.}
  \label{fig1}
\end{figure}

The calculated energy levels for $^{66}$As are compared with the
experimental data \cite{Grzywacz2} in Fig. \ref{fig1}. As one can
see, our calculation produces a level scheme similar to the
observed one \cite{Grzywacz2}. We correctly obtain the $T=1$
ground state band (the $0^+$-$2^+$-$4^+$ sequence) and the $1_1^+$
state as the lowest state of $T=0$. We notice that in $^{66}$As,
the experimental $T=0$, $1^+$ state is marked as uncertain
\cite{Grzywacz2}, and in the heavier odd-odd $N=Z$ nuclei
$^{70}$Br \cite{Jenkins0} and $^{74}$Rb \cite{OLeary}, such a
$1^+$ state is experimentally missing. Above the $1_1^+$ state,
there are the first $J=3$ state $3_1^+$ and the first $J=5$ state
$5_1^+$. Interestingly enough, the calculated $E2$ transition
strength $B(E2:3_1^+ \rightarrow 1_1^+)$ is only 1.3 Weisskopf
unit (W.u.), which is much smaller than the value 16 W.u. for
$B(E2:3_2^+ \rightarrow 1_1^+)$. The calculated $B(E2:5_1^+
\rightarrow 3_2^+)$ is also smaller than $B(E2:5_2^+ \rightarrow
3_2^+)$. We are thus compelled to conclude that the $3_2^+$ and
$5_2^+$ states belong to the same collective sequence with the
$1_1^+$ state, which resembles the observed band of $T=0$
\cite{Grzywacz2}, but the $3_1^+$ and $5_1^+$ states do not.
Whereas our calculated $5_1^+$ state may correspond to the
observed $5^+$ isomeric state \cite{Grzywacz1} (see discussions
below), the predicted $3_1^+$ state seems to be missing in the
experiment.

In order to understand the structure of these states, we study the
wave-functions by calculating expectation values of proton and
neutron number operators in the four orbitals, as well as those of
spin and isospin for nucleons from the subspaces
$pf=(p_{3/2},f_{5/2},p_{1/2})$ and $g_{9/2}$. To facilitate the
discussion, we denote the former quantity by ($\langle n_a^\pi
\rangle $, $\langle n_a^\nu \rangle $), and the latter by ($J_i$,
$T_i$), with $i=pf$ or $g_{9/2}$. The expectation values of spin
and isospin, $J_i$ and $T_i$, are respectively evaluated from
$J_i=[\langle ({\hat j}_i)^2 \rangle +1/4]^{1/2} -1/2$ and
$T_i=[\langle ({\hat t}_i)^2 \rangle +1/4]^{1/2} -1/2$, where
${\hat j}_i$ is the spin operator and ${\hat t}_i$ the isospin
operator. Table \ref{table1} shows these expectation values and
the calculated spectroscopic $Q$-moment, defined as
$Q=\sqrt{16\pi/5} \sum_{\tau=\pi,\nu} e_{\rm eff}^\tau \langle
r^2Y_{20} \rangle_\tau$.

\begin{table}[b]
\caption{Expectation values of nucleon numbers in the four
orbitals,
         those of spin and isospin of nucleons in the subspaces
         $pf=(p_{3/2},f_{5/2},p_{1/2})$ and $g_{9/2}$,
         and calculated spectroscopic $Q$-moments (in $e$ fm$^2$),
         for the $T=0$ states in $^{66}$As.}
\begin{tabular}{c|cccc|cccc|c}   \hline
        & \multicolumn{4}{c}{$\langle n_a^\pi \rangle  =\langle n_a^\nu \rangle $}
        & \multicolumn{4}{|c|}{$J_i$, $T_i$}
        & \\ \hline
 $T=0$ & $p_{3/2}$ & $f_{5/2}$ & $p_{1/2}$ & $g_{9/2}$
        & $J_{pf}$ & $T_{pf}$ & $J_{g9/2}$ & $T_{g9/2}$ & $Q$  \\ \hline
 $1_1^+$  & 2.13 & 2.16 & 0.52 & 0.19 & 1.30 & 0.22 & 0.67 & 0.22 & -21.0 \\
 $3_2^+$  & 2.07 & 2.23 & 0.54 & 0.16 & 3.06 & 0.19 & 0.64 & 0.19 & -26.3 \\
 $5_2^+$  & 2.04 & 2.14 & 0.67 & 0.15 & 4.97 & 0.18 & 0.64 & 0.18 & -30.9 \\
 $7_1^+$  & 1.97 & 2.36 & 0.53 & 0.14 & 6.90 & 0.17 & 0.74 & 0.17 & -28.8 \\ \hline
 $3_1^+$  & 2.12 & 2.19 & 0.52 & 0.18 & 2.96 & 0.22 & 0.72 & 0.22 & +46.2 \\
 $5_1^+$  & 1.99 & 2.33 & 0.55 & 0.13 & 4.97 & 0.17 & 0.58 & 0.17 & -15.0 \\ \hline
 $9_1^+$  & 1.59 & 1.82 & 0.55 & 1.04 & 1.77 & 0.07 & 8.93 & 0.07 & -84.7 \\
$11_1^+$  & 1.61 & 1.75 & 0.61 & 1.03 & 2.74 & 0.06 & 8.95 & 0.06
& -90.1 \\ \hline
 $9_2^+$  & 1.63 & 1.66 & 0.58 & 1.14 & 2.77 & 0.90 & 7.77 & 0.81 & -69.1 \\ \hline
\end{tabular}
\label{table1}
\end{table}

From Table \ref{table1}, it is interesting to observe that the
sign of spectroscopic $Q$-moment in $3_1^+$ state is opposite to
those of all other states. Hence the shape of the $3_1^+$ state is
predicted to be oblate, in contrast to the prolate shape for other
states. This conclusion has also been confirmed by the potential
energy surface for $^{66}$As with the same method as used in Ref.
\cite{Kaneko}. In addition to the rather small $B(E2:3_1^+
\rightarrow 1_1^+)$, the calculated energy difference between the
$3_1^+$ and the lowest $T=0$ state $1_1^+$ is only about 0.07 MeV.
Thus the present model calculation predicts a long-lived $3^+$
shape isomer.

An isomeric $3^+$ state has been found in the lighter odd-odd
$N=Z$ nucleus $^{62}$Ga \cite{Vincent,Rudolph}. According to Ref.
\cite{Rudolph}, the isomeric nature of this state simply arises
from the small transition energy. The measured half-life is
$T_{1/2}=4.9(^{14}_{13})$ ns and the corresponding $B(E2:3_1^+
\rightarrow 1_1^+)$ value is $\sim 13$ W.u. To compare with these
data in $^{62}$Ga, we have carried out a shell model calculation.
Our model reproduces considerably well the observed energy levels
in $^{62}$Ga \cite{Rudolph}. The $B(E2:3_1^+ \rightarrow 1_1^+)$
value is calculated to be 8.7 W.u., which is comparable to those
from other shell model calculations \cite{Vincent,Rudolph}.
However, our calculation suggests that the isomeric $3^+$ state in
$^{62}$Ga belongs to the same $T=0$ band with the $1^+$ state as
the bandhead, and the calculated spectroscopic $Q$-moment for the
$3^+$ state has the same sign as the other states.

Thus, unlike the $3^+$ isomer in $^{62}$Ga, the lowest $T=0$,
$3^+$ state in $^{66}$As is a shape isomer in nature. This
provides an example in odd-odd $N=Z$ nuclei that a low-lying state
corresponding to oblate shape coexists with those having prolate
shape. Shape coexistence phenomenon, i.e. occurrence of two or
more stable shapes in a nucleus at comparable excitation energies,
has been known in the neighboring even-even $N=Z$ nuclei. In
$^{72}$Kr, the excited $0^+$ bandhead state at 0.671 MeV was
identified as a shape isomer with lifetime 38 ns \cite{Bouchez}. A
similar shape isomer in $^{68}$Se has been predicted
\cite{Sun,Sun3} to exist, awaiting experimental confirmation.
Thus, our results here reinforce the early claim \cite{Lister}
about violent shape change in this mass region. We may further
speculate that the non-observation of the anticipated $T=0$
bandhead $1^+$ state in $^{70}$Br and $^{74}$Rb may imply a highly
mixed structure in this state due to the shape coexistence nature,
thus suppressing the decay.

In Ref. \cite{Grzywacz1}, an isomeric state $5_1^+$ at 1.357 MeV
with half-life $T_{1/2}=1.1(1)$ $\mu$s was reported, which was
interpreted based on a non-collective picture. Our calculation
reproduces a decay scheme similar to the observed one. However,
the calculated value 7.4 W.u. for $B(E2:5_1^+ \rightarrow 3_2^+)$
is about twenty times larger than the experimental value 0.34(9)
W.u. Thus our calculation can not explain the isomerism of the
experimental $5_1^+$ state. The nucleon occupation numbers
$\langle n_a \rangle$ in Table \ref{table1} show that all the
low-lying states below the $9_1^+$ one contain a strong
configuration mixing in the $pf$ subspace. In this sense, the
$5_1^+$ state is regarded as a collective state. Thus in our
picture, a long half-life of the $5_1^+$ state should be
attributed to its different collective structure from other
lower-lying states to which it decays. Our calculated $Q$-moments
suggest that the $5_1^+$ state has indeed a different structure
from the lower-lying states of the $T=0$. However, the difference
seems to be insufficient to explain the $5_1^+$ isomerism found in
experiment. We leave this as an open question.

Now we move our discussion to the $J^\pi=9^+$ isomeric state
discovered in Refs. \cite{Grzywacz1,Grzywacz2}. This state has a
long half-life $T_{1/2}=8.2(5)$ $\mu$s, and the measured $E2$
transition to the $7^+$ state is only 0.044(6) W.u.
\cite{Grzywacz1}. Our calculated $B(E2)$ value for this transition
is 0.023 W.u., which reproduces well the observed one. To see what
causes this state to have such a long lifetime, let us study the
quantities listed in Table \ref{table1}. The calculated
spectroscopic $Q$-moment for the $9_1^+$ state shows a very
different value from the other lower-lying states, indicating a
different structure. The change in structure in the $9_1^+$ state
is caused by rotational alignment of $g_{9/2}$ particles. From
Table \ref{table1}, it can be clearly seen that $\langle
n_{g9/2}^\pi \rangle =\langle n_{g9/2}^\nu \rangle \approx 1$,
$J_{g9/2} \approx 9$ and $T_{g9/2} \approx 0$ for the states
$9_1^+$ and $11_1^+$, indicating a complete $g_{9/2}$
one-proton-one-neutron ($1p1n$) alignment. The $T=0$, $1p1n$
alignment in even-even nuclei has been discussed in Refs.
\cite{Hase1,Hase2}. When comparing the occupation numbers $\langle
n_a \rangle $ in $^{64}$Ge and $^{66}$As, we find that a state
with a pair alignment in $^{66}$As may be symbolically denoted as
$^{64}$Ge$\otimes (g_{9/2}^\pi g_{9/2}^\nu)$. In the language of
the projected shell model \cite{Hara}, this is a two-quasiparticle
configuration. In contrast, the states below the $9_1^+$ state
comprise nucleons mainly from the $pf$ shell. It is then clear
that the electromagnetic transitions to a state below $9_1^+$ is
strongly retarded by single-particle operators. Thus, the isomeric
nature of the $9_1^+$ state is attributed to the fact that the
$9_1^+$ state abruptly changes the structure by breaking a $T=0$,
$1p1n$ pair from the $g_{9/2}$ orbitals and aligning them with the
rotational axis.

The above analysis actually suggests a new class of isomers, which
may occur in $N\approx Z$ systems only. An isomeric state may be
formed at a high spin when a rotational alignment of a pair is
completed in this state. The involved pair consists of one proton
and one neutron from the high-$j$ intruder orbitals ($g_{9/2}$ in
our example here), which have distinct properties from the nearby
normal parity orbitals. The significant structure change in the
wavefunctions before and after the alignment suppresses
electromagnetic transitions, resulting in a long-lived state. This
conclusion is robust because even if we switch off the monopole
corrections in Eq. (\ref{eq:1}), the $B(E2:9_1^+ \rightarrow
7_1^+)$ value still remains very small. The suppression of
transitions is not due to existence of an energy barrier or
violation of the selection rules. It is thus obvious that this
type of isomer does not belong to any of the three isomer classes
already discussed in the literature \cite{Nature}. We may term it
{\it j-isomer} since the underlying mechanism leading to it is the
presence of significant components of the high-$j$ intruder states
in the wavefunctions. The observed 8.2 $\mu$s, $9^+$ state in
Refs. \cite{Grzywacz1,Grzywacz2} is an example of a $j$-isomer.

\begin{figure}[b]
\includegraphics[width=8cm,height=7.5cm]{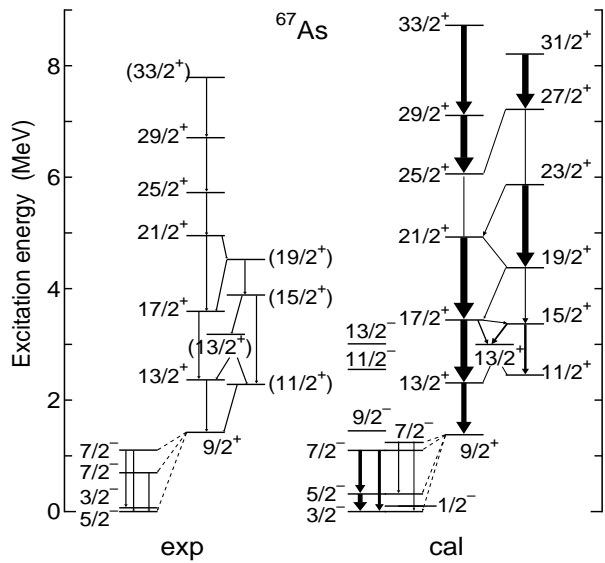}
  \caption{Experimental and calculated energy levels in $^{67}$As.
           Widths of the arrows between the calculated states (cal)
           denote relative values of the calculated $B(E2)$ (solid lines) and
           $B(E3)$ (dotted lines).}
  \label{fig2}
\end{figure}

In our calculation, there is the second $9^+$ state lying slightly
above 4 MeV. As can be seen from Table \ref{table1}, the structure
of this state is again very different from the others. The
wavefunction of this $9_2^+$ state has a large component of a
$T=1$, $g_{9/2}^\pi g_{9/2}^\nu$ aligned pair with $J\approx 8$,
and the core excluding the $g_{9/2}$ nucleons has $J \approx 2.8$
and $T=0.9$. The calculated $B(E2)$ values to the lower-lying
states $7_1^+$, $9_1^+$, and $11_1^+$ do not exceed 0.7 W.u.
Therefore, the $9_2^+$ state possibly has a long lifetime.

The odd-proton $N=Z+1$ nucleus $^{67}$As has a known isomeric
state ${9/2}^+$ at 1.422 MeV with $T_{1/2}=12(2)$ ns. To study the
structure of this state, we have performed shell model
calculations for $^{67}$As. Energy levels obtained by the
calculation are compared with the experimental ones \cite{Jenkins}
in Fig. \ref{fig2}. Again our calculation reproduces well the
experimental level scheme. The only disagreement with data is the
ground state spin. While data suggested that the ground state has
the spin $I^\pi={5/2}^-$, our calculation obtained a ${3/2}^-$
state as the lowest state in $^{67}$As. Repeated calculations by
changing the parameters within the extended $P+QQ$ Hamiltonian
have not succeeded to get a ${5/2}^-$ ground state. We notice that
$^{65}$Ge has the ground state ${3/2}^-$, which is correctly
reproduced by our calculations. Presently, we can not explain why
$^{65}$Ge with 33 neutrons has the ground state ${3/2}^-$, while
$^{67}$As with 33 protons should have the ground state ${5/2}^-$.

\begin{table}[b]
\caption{Expectation values of proton and neutron numbers
         in the four orbitals, calculated for those low-lying states in $^{67}$As.
         Calculated spectroscopic $Q$-moments (in $e$ fm$^2$) are also tabulated.}
\begin{tabular}{c|cccc|cccc|c}   \hline
        & \multicolumn{4}{c}{$\langle n_a^\pi \rangle $}
        & \multicolumn{4}{|c|}{$\langle n_a^\nu \rangle $} & \\ \hline
$^{67}$As & $p_{3/2}$ & $f_{5/2}$ & $p_{1/2}$ & $g_{9/2}$
         & $p_{3/2}$ & $f_{5/2}$ & $p_{1/2}$ & $g_{9/2}$ & $Q$  \\ \hline
 $1/2_1^-$ & 1.99 & 2.15 & 0.72 & 0.15 & 2.33 & 2.30 & 1.19 & 0.17 &   1.1 \\
 $3/2_1^-$ & 2.17 & 2.12 & 0.54 & 0.17 & 2.48 & 2.60 & 0.73 & 0.19 &  23.9 \\
 $5/2_1^-$ & 2.06 & 2.13 & 0.68 & 0.13 & 2.36 & 2.74 & 0.74 & 0.17 & -13.8 \\
 $7/2_1^-$ & 2.04 & 2.19 & 0.62 & 0.15 & 2.36 & 2.78 & 0.68 & 0.18 & -24.2 \\
 $7/2_2^-$ & 2.16 & 1.66 & 1.00 & 0.18 & 2.33 & 2.35 & 1.12 & 0.20 &  -1.5 \\ \hline
 $9/2_1^+$ & 1.69 & 1.93 & 0.65 & 0.73 & 2.26 & 2.43 & 0.81 & 0.51 & -57.9 \\
$13/2_1^+$ & 1.78 & 1.88 & 0.62 & 0.72 & 2.22 & 2.62 & 0.63 & 0.53 & -57.8 \\
$11/2_1^+$ & 1.75 & 1.77 & 0.75 & 0.73 & 2.19 & 2.34 & 0.88 & 0.59 & -51.6 \\
$15/2_1^+$ & 1.85 & 1.90 & 0.61 & 0.64 & 2.14 & 2.52 & 0.66 & 0.69
& -56.4 \\ \hline
\end{tabular}
\label{table2}
\end{table}

Further inspection of the wavefunctions shows that, similar to its
even-even and odd-odd neighbors, the low-lying states in this
odd-proton nucleus are also characterized by shape coexistence. We
have calculated for $^{67}$As the expectation values $\langle
n_a^\pi \rangle $ and $\langle n_a^\nu \rangle $, as well as $J_i$
and $T_i$. Part of the results is shown in Table \ref{table2}. The
numbers indicate a strong configuration mixing in the $pf$
subspace as found in $^{66}$As. Protons and neutrons appear to be
excited from the $p_{3/2}$ to the $f_{5/2}$ orbital in the
${5/2}_1^-$, ${3/2}_1^-$, and ${7/2}_1^-$ states. This effect
lowers the ${3/2}_1^-$ state in the calculation. The ${5/2}_1^-$
and ${3/2}_1^-$ states are not simply single-particle-like states.
The calculated spectroscopic $Q$-moment suggests an oblate shape
for ${3/2}^-$ but a prolate shape for ${5/2}^-$. In addition, the
present model predicts a very low-lying ${1/2}^-$ state. The
${1/2}^-$ state actually becomes the ground state in the
neighboring nucleus $^{67}$Ge, which is correctly reproduced by
our model \cite{Hase2}. We can thus conclude that our model
reproduces considerably well the low-lying negative-parity and the
positive-parity states.

The long life of the ${9/2}_1^+$ state corresponds to suppressed
electromagnetic transitions to the lower negative-parity states.
Calculated reduced $E3$ transition strengths from ${9/2}_1^+$ to
the ${5/2}_1^-$, ${3/2}_1^-$, ${7/2}_1^-$, and ${7/2}_2^-$ states
are 0.057, 0.074, 0.001, and 0.010 W.u., respectively. These
rather small $B(E3)$ values are consistent with the long life of
the ${9/2}_1^+$ state. Naively, one may think of a simple picture
for the ${9/2}_1^+$ state that the last proton occupies the
$g_{9/2}$ orbital. However, the proton and neutron occupation
numbers in Table \ref{table2} indicate a completely different
structure. The neutron occupation number $\langle n_a^\nu \rangle
$ in the $g_{9/2}$ orbital is unexpectedly large. The expectation
values of spin and isospin of nucleons in the $g_{9/2}$ orbital,
$J_{g9/2}$ and $T_{g9/2}$, are 4.54 and 0.53, respectively. These
values, $\langle n_a^\pi \rangle =0.73$, $\langle n_a^\nu \rangle
=0.51$, $J_{g9/2}\approx 9/2$, and $T_{g9/2}\approx 1/2$, suggest
two dominant configurations: one has a proton (and probably a
fraction of $J=0$ neuron pair) in the $g_{9/2}$ orbital and the
other has a neutron (and probably a fraction of $J=0$ proton pair)
in the $g_{9/2}$ orbital. The large components of the
configurations may be symbolically expressed as a mixture of
$^{66}$Ge$\otimes g_{9/2}^\pi$ and $^{66}$As$\otimes g_{9/2}^\nu$.
To have these configurations, the neutron $g_{9/2}$ and the proton
$g_{9/2}$ orbitals must be occupied simultaneously. Simultaneous
occupation of the same orbital by protons and neutrons is
certainly the unique property for $N\approx Z$ nuclei. On the
other hand, the negative-parity states below the ${9/2}_1^+$ state
are collective states with strongly mixed configurations in the
$pf$ subspace. The electromagnetic transitions are therefore
strongly hindered by single-particle operators.

In conclusion, to study the structure of isomeric states, we have
calculated the odd-odd $N=Z$ nucleus $^{66}$As and the odd-proton
$N=Z+1$ nucleus $^{67}$As, in the framework of the spherical shell
model. By using the extended $P+QQ$ Hamiltonian, we are able to
reproduce the level schemes for these two isotopes. In $^{66}$As,
a $T=0$, $3^+$ state has been predicted at about 1 MeV. We suggest
that this is a shape isomer, and once confirmed, it may represent
the first example of a shape isomer in an odd-odd nucleus in this
mass region. The occurrence of a shape isomer in $^{66}$As
supports the picture of prolate-oblate shape coexistence at
low-excitations, and may provide a clue to the puzzle of
non-observation of the anticipated $T=0$ bandhead state in some
heavier odd-odd $N=Z$ nuclei. The structure of the experimentally
known isomeric state $9^+$ in $^{66}$As and $9/2^+$ in $^{67}$As
have been studied and termed $j$-isomers, which, unlike the three
well-known classes of isomers, are unique to $N\approx Z$ nuclei.
The formation of $j$-isomer requires either an alignment of a
neutron-proton pair from the high-$j$ intruder orbitals (in
$^{66}$As), or a simultaneous occupation of these neutron and
proton high-$j$ orbitals (in $^{67}$As).

To go to heavier, deformed mass region with $A>70$, one may refer
to shell models based on a deformed basis, such as the projected
shell model \cite{Hara}. From the theoretical point of view, it
would be interesting to compare, while employing the same
Hamiltonian, the results of both types of shell models constructed
in a spherical or in a deformed basis. The nuclei discussed here
are perfect cases for a comparison. Work along these lines is in
progress.

Y.S. acknowledges communications with G. de Angelis at the early
stage of the present work. This work is partly supported by NSF
under contract PHY-0140324, and by the Grant-in-Aid of the
Promotion and Mutual Aid Corporation for Private Universities of
Japan in 2003 and 2004.



\end{document}